\documentclass[twocolumn,prb,showpacs,amsmath,amssymb]{revtex4}

\usepackage{graphicx}
\usepackage{dcolumn}
\usepackage{bm}
\def\cm{cm$^{-1}$}

\begin{document}

\title{
Low-Frequency Optical Properties of $\beta^{\prime\prime}$-(BEDO-
TTF)$_5$[CsHg(SCN)$_4$]$_2$:\\
Indications of Electronic Correlations in a 1/5-Filled Two-Dimensional
Conductor}

\author{N. Drichko$^{1,2}$}
\author{K. Petukhov$^1$}
\author{M. Dressel$^1$}
\author{O. Bogdanova$^3$}
\author{E. Zhilyaeva$^3$}
\author{R. Lyubovskaya$^3$}
\author{A. Greco$^4$}
\author{J. Merino$^5$}
\affiliation{
$^1$ 1.~Physikalisches Institut, Universit{\"a}t Stuttgart, Pfaffenwaldring 57,
70550 Stuttgart, Germany \\
$^2$ Ioffe Physico-Technical Institute, 194021 St.Petersburg, Russia\\
$^3$ Institute of Problems of Chemical Physics, Russian Academy of Sciences,
Chernogolovka, Russia\\
$^4$ Facultad de Ciencias Exactas Ingenier\'ia y Agrimensura e Instituto de
F\'isica Rosario (UNR-CONICET), Rosario, Argentina\\
$^5$ Max-Planck-Institut f\"ur Festk{\"o}rperforschung, D-70506 Stuttgart,
Germany}
\date{\today}
\begin{abstract}
The polarized reflectivity of $\beta^{\prime\prime}$-(BEDO-TTF)$_5$[CsHg(SCN)$_4$]$_2$ is studied in the
infrared range between 60~\cm\ and 6000~\cm\ from room temperature down to 10~K. Already at $T=300$~K a
pseudogap in the optical conductivity is present of about 300~\cm; the corresponding maximum in the
spectrum shifts to lower frequencies as the temperature decreases.  In contrast to quarter-filled
BEDT-TTF-based conductors of the $\beta^{\prime\prime}$-phase a robust Drude component in the conductivity
spectra is observed which we ascribe to the larger fraction of charge carriers associated with the
1/5-filling of the conduction band. This observation is corroborated by exact diagonalization calculations
on an extended Hubbard model on a square lattice for different fillings. A broad band at 4000~\cm\ which
appears for the electric field polarized parallel to the stacks of the BEDO-TTF molecules is associated to
structural modulations in the stacks; these modulations lead to a rise of the dc and microwave resistivity
in 100 to 30~K temperature range.
\end{abstract}


\maketitle

\section{Introduction}

Molecular quasi two-dimensional organic conductors based on
BEDT-TTF (where BEDT-TTF stands for
bis\-ethylene\-di\-thio\--tetra\-thia\-fulvalene) and its
derivatives are known to be excellent model objects to study the
effects of electron-electron correlations and charge ordering of
different nature in two dimensional (2D)
systems.\cite{Ishiguro98,Jerome94} According to theoretical
studies,\cite{McKenzie00b,Kino95,Seo00,McKenzie01,Georges96,Calandra02}
the electronic parameters which are crucial for the metallic
properties and formation of charge ordering (CO) are the filling
of the conduction band and the size of the electronic correlation
energy compared to the width of conductance band or transfer
integrals. In BEDT-TTF-based compounds these physical parameters
can be tuned chemically by changing the structure or/and chemical
composition of the crystals. The width of the conduction band
increases with the orbital overlap of neighboring molecules. The
filling of the conduction band depends upon both the anionic
charge and upon the ratio of cations and anions. Recently a large
effort has been devoted to systematically control the band filling
and to study its influence on the physical
properties.\cite{Mori02,Mori} However, most of 2D organic
conductors have quarter-filled bands (charge $+0.5e$ per
molecule), or effectively half-filled bands ($+1e$ per dimer) for
dimerized structure. The isostructural family
$\beta^{\prime\prime}$-(BEDO-TTF)$_5$\-[$M$Hg\-(SCN)$_4$]$_2$ ($M$
= K, Rb, Cs, NH$_4$, Li) \cite{zhilya,Lyubovskii} was of interest
to us, as these salts crystallize in a 5:2 ratio of cations and
anions leading to a 1/5 filling of the conduction band.

It is important for our study, that when going from BEDT-TTF to
BEDO-TTF (bis-ethylene\-di\-oxy\--tetra\-thia\-fulvalene) the
on-site electron-electron correlation energy does not change
considerably.\cite{saito} However, the probability to obtain an
insulating and superconducting salt among the BEDO-TTF family is
lower than among BEDT-TTF based ones. Horiuchi et al.\ suggested
that these differences are due to the fact that the BEDO-TTF
molecules tend to form conducting layers which are more
two-dimensional.\cite{horiuchi} Indeed, within the
(BEDO-TTF)$_5$\-[$M$Hg\-(SCN)$_4$]$_2$ ($M$ = K, Rb, Cs, NH$_4$,
Li) family most members exhibit a metallic behavior in the
highly-conducting plane down to low temperatures. The only
exception is
$\beta^{\prime\prime}$-(BEDO-TTF)$_5$\-[CsHg\-(SCN)$_4$]$_2$ which
shows non-metallic behavior of the dc conductivity in one direction
of the plane, while in perpendicular direction conductivity
remains metallic down to 4~K. Previous optical studies of
$\beta^{\prime\prime}$-(BEDO-TTF)$_5$\-[CsHg\-(SCN)$_4$]$_2$ in
the 700-6000~\cm\ region have also revealed an anisotropic
temperature behavior of the optical conductivity;\cite{we} the
reason for this unusual behavior might be a dimerization in the
stacks of BEDO-TTF molecules. These investigations also led to the
conclusion that at frequencies below the mid-infrared a non-Drude
response has to be expected.

Optical investigations and theoretical studies of BEDT-TTF based
materials with a metallic ground state have shown that the effects
of electronic correlations lead to deviations from the simple Drude
behavior; however, the results strongly depend on the band
filling. For the metallic half-filled $\kappa$-phase compounds a
narrow Drude-peak was observed only at temperatures below
50~K,\cite{Kornelsen89} in accordance with dynamical
mean-field-theory calculations \cite{Georges96} on a frustrated
lattice at half-filling with strong on-site Coulomb repulsion $U
\approx W$ ($W$ being the bandwidth).\cite{Georges96,McKenzie00b}
Quarter-filled systems, on the other hand, remain metallic for any
$U$,\cite{Kino95} as long as the nearest-neighbor interaction $V$
is negligible; by increasing $V$ beyond some critical value $V_c$,
charge-ordering phenomena become
relevant.\cite{Seo00,McKenzie01,Calandra02} In fact this was
recently demonstrated in the spectra of BEDT-TTF-based organic
conductors with an $\alpha$-structure of the conducting
layers.\cite{Dressel03} While
$\alpha$-(BEDT-TTF)$_2$\-NH$_4$Hg(SCN)$_4$ shows metallic
properties at any temperature, in
$\alpha$-(BEDT-TTF)$_2$\-KHg(SCN)$_4$ a gap opens at about
200~\cm\ at temperatures below 200~K with a narrow Drude component
remaining. However, in some metallic $\beta^{\prime\prime}$-packed
BEDT-TTF salts which are quarter filled, no Drude peak was
observed down to 10~K.\cite{dong1999}

This has motivated us to extend the optical measurements of
$\beta^{\prime\prime}$-(BEDO-TTF)$_5$\-[CsHg\-(SCN)$_4$]$_2$ to
frequencies well below 700~\cm. We got an opportunity to study the
expected deviations from a simple Drude behavior for this 1/5
filled system and to compare them to theoretical expectations and
to the low-frequency behavior of a $\beta^{\prime\prime}$-packed
BEDT-TTF based 1/4 filled system.\cite{dong1999}  In this first
low-frequency optical experiment on a BEDO-TTF salt, we have also
investigated the temperature dependence of the anisotropy which
characterizes the in-plane conductivity.

\section{Experimental}
Single crystals of
$\beta^{\prime\prime}$-(BEDO-TTF)$_5$[CsHg(SCN)$_4$]$_2$ were
prepared by electrochemical oxidation of BEDT-TTF as described
in Ref.~[\onlinecite{zhilya}]. In structure of these crystals the BEDO-TTF
mole\-cules form conducting layers in the $ab$-plane and alternate
with insulating anion layers along $c$-axis.
$\beta^{\prime\prime}$-phase means that the molecules form stacks
parallel to the $2a$-$b$ direction; along the stacks the overlap
integrals show some modulation. The overlap integrals to the
molecules in the neighboring stacks are nearly two times larger
than between the molecules along the stacking direction.\cite{Lyubovskii}
According to
the stoichiometry, each BEDO-TTF molecule carries an average
charge of $+0.4e$, thus the conduction band is 1/5 filled. The
crystals grow as black parallelogram-shaped plates, some of them
reached $2\times 1$~mm$^2$ in size. The largest extension of the
crystals is parallel to the stacking direction of the BEDO-TTF
molecules in the conducting layer, i.e.\ the $2a$-$b$ direction.

\begin{figure}
\includegraphics*[width=6cm]{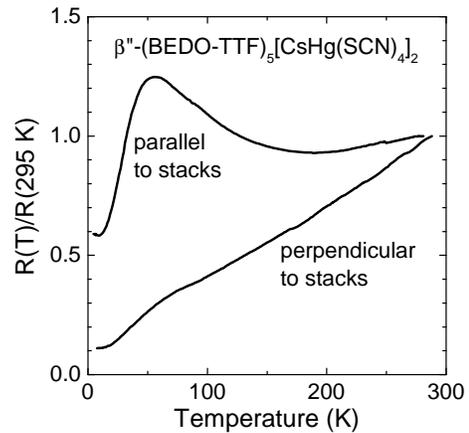}
\caption{\label{dc} Temperature dependence of the normalized dc resistance of
$\beta^{\prime\prime}$-(BEDO-TTF)$_5$[CsHg(SCN)$_4$]$_2$ parallel and
perpendicular
to the stacking direction.}
\end{figure}
The $\beta^{\prime\prime}$-(BEDO-TTF)$_5$[CsHg(SCN)$_4$]$_2$
crystals were previously characterized by standard dc transport
measurements parallel and perpendicular to the stacks.\cite{we}
The values of the room temperature resistivity for the two
directions are around 0.2 to 0.5~$\Omega$cm. The temperature
dependence of the dc resistivity is plotted in Fig.~\ref{dc}.
Perpendicular to the stacking direction the resistivity remains
metallic down to helium temperatures; parallel to the stacks the
resistivity changes only little between 270~K and 50~K indicating
a semiconducting behavior, and becomes metallic below 50~K.

In order to investigate the anisotropic transport properties
without applying contacts, we performed microwave
re\-sis\-ti\-vity measur\-ements by means of cavity perturbation
technique;\cite{1,2,3} the experimental setup and measurement
technique are discussed in detail elsewhere.\cite{Petukhov03} For
the experiments reported here, the sample was fixed to a quartz
rod positioned in the electric field maximum of a cylindrical
copper resonator which operates in the TE$_{011}$ transmission
mode at a resonance frequency of 24~GHz. At any temperature down
to 1.5~K the sample could be rotated inside the cavity during the
measurement, allowing to determine its
anisotropy {\it in situ}.\cite{Petukhov03}

\begin{figure}
\includegraphics*[width=6cm]{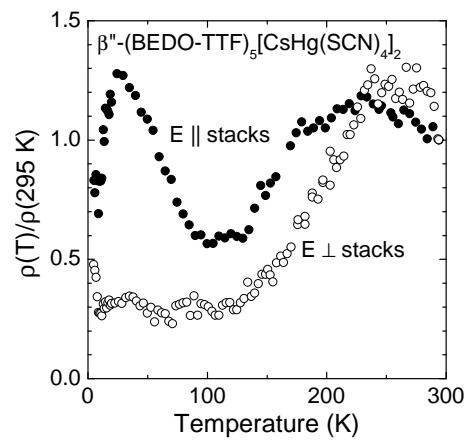}
\caption{\label{microwave}
Temperature dependance microwave resistivity of
$\beta^{\prime\prime}$-(BEDO-TTF)$_5$[CsHg(SCN)$_4$]$_2$ measured at
24~GHz  with the electric field parallel (full symbols) and perpendicular (open
circles)
to the stacks of the BEDO-TTF molecules.}
\end{figure}
In Fig.~\ref{microwave} the normalized microwave resistivity of
$\beta^{\prime\prime}$-(BEDO-TTF)$_5$[CsHg(SCN)$_4$]$_2$ measured
parallel and perpendicular to the stacks is plotted as a function
of temperature. Perpendicular to the direction of the stacks the
resistivity $\rho(T)$ exhibits a metallic behavior between 300 and
150~K, with only little change on cooling below 150~K. Along the
stacks we find a decrease in $\rho(T)$ when going down to 100~K,
followed by some insulating behavior; below 50~K a metallic regime
is entered again.\cite{footnote} These results are consistent
with dc resistivity measurements  displayed in
Fig.~\ref{dc}.

\begin{figure*}
\includegraphics*[width=15cm]{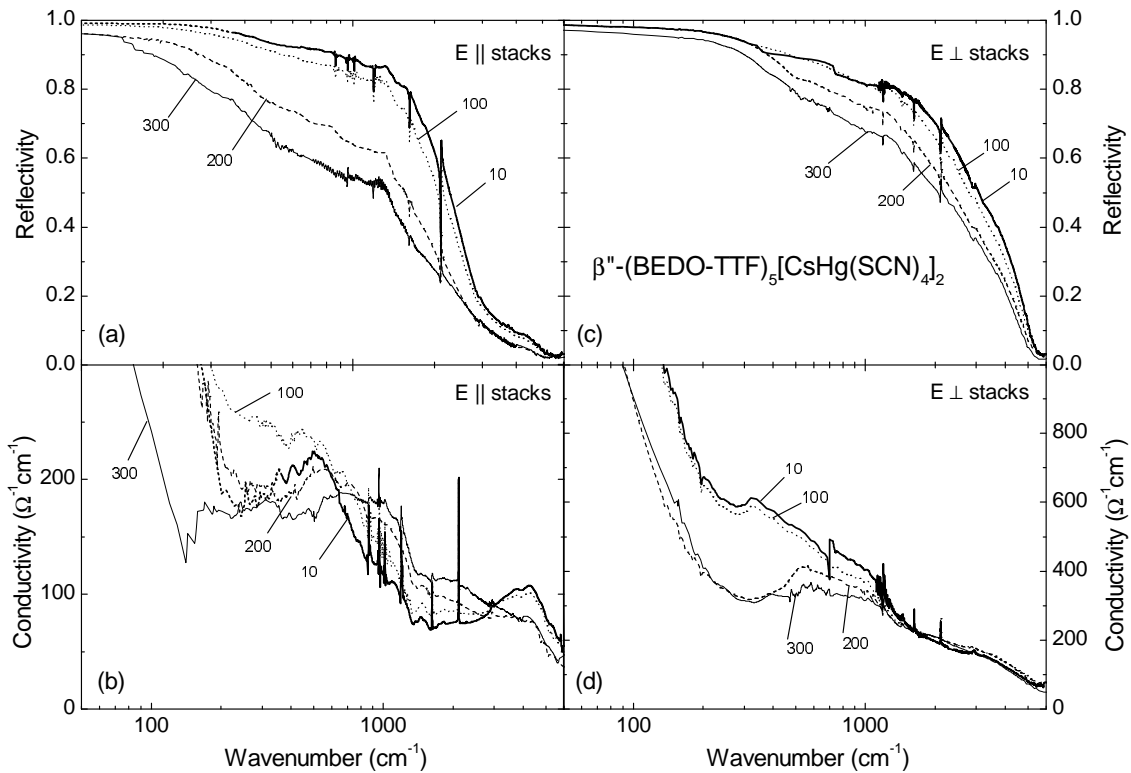}
\caption{\label{optics}
Reflectivity and optical conductivity spectra of
$\beta^{\prime\prime}$-(BEDO-TTF)$_5$[CsHg(SCN)$_4$]$_2$
for polarizations E$\parallel$stacks and
E$\bot$stacks
at 300, 200, 100 and 10~K.
}
\end{figure*}
The polarized reflectivity R($\omega$) spectra of the two main
directions in the conducting plane of
$\beta^{\prime\prime}$-(BEDO-TTF)$_5$[CsHg(SCN)$_4$]$_2$ are
measured in the frequency range of 60-6000~\cm\ using a modified
Buker IFS113v Fourier transform spectrometer equipped with three
different light sources and various radiation detectors, including
a helium cooled bolometer. As reference we use an aluminum mirror
which replaces the sample by translation. With the help of proper
polarizers the optical response was measured for the electric field
$E$ polarized parallel and perpendicular to the stacks of the
single crystal. The sample is cooled
down to 10~K in a Cryo\-Vac helium bath cryostat with a very slow
cooling rate of 1~K/min. The overall behavior of the reflectivity
is in good agreement with the previous results obtained in the
spectral region of 700-6000~\cm\ which were measured by a
microreflectance technique.\cite{we} However due to small defects
on the  surface of the single crystals, our values of reflectivity
are slightly lower. The correct absolute values are obtained by
normalizing the reflectivity with a constant factor in order to
receive perfect agreement with the microscopic measurements at
room temperature. Since the low-temperature spectra $E$ parallel
to the stacks  are very noisy below 300~\cm, for clarity reasons
we show a smoothed curve as a dashed line in Fig.~\ref{optics}.

\section{Results and Analysis}
The spectra of the reflectivity R($\omega$) and the optical
conductivity $\sigma(\omega)$ in the two polarizations parallel
and perpendicular to the stacks are presented in
Fig.~\ref{optics}. For the electric field $E$ of the incident
light parallel to the stacks, the room-temperature reflectivity is
lower than for the perpendicular polarization, in agreement with
the smaller overlap of the BEDO-TTF orbitals in this direction.
The reflectivity $R(\omega)$ increases to lower frequencies and
the plasma edge is located around 5000~\cm. In the frequency range
from 700 to 1600~\cm\ weak vibrational features are seen, their
detailed assignment to $a_g$ vibrations activated by
electronic-molecular vibrational (EMV) coupling is published in
Ref.~[\onlinecite{we}]. With lowering temperature, $R(\omega)$ grows in
both polarizations of the electric field, however in different
ways. For $E$ normal to the stacks the reflectivity in the frequency range
from 60 to 700~\cm\ increases upon cooling down to 150~K, but
remains unchanged below this temperature. Parallel to the stacks
the reflectivity in the 60-700~\cm\ range changes rapidly between
200 and 150~K, and continues to grow slightly below this temperature.
There is
basically no change of $R(\omega)$ on cooling between 50 and 10~K.
At low temperatures the reflectivity along the stacks reaches the
corresponding $R(\omega)$ values of $E$ perpendicular to the stacks.
With other words, the large
room-temperature anisotropy of reflectivity in the conducting
plane observed in the spectral range from 60 to 700~\cm\
disappears at temperature $T\leq 50$~K.

The optical conductivity spectra are calculated with the help of
the Kramers-Kronig transformation. The measured $R(\omega)$
spectra are extrapolated by the Hagen- Rubens formula for
$\omega\rightarrow 0$; above 6000~cm$^{-1}$ the results of our
room-temperature measurements \cite{Drichko_rus} and extrapolation
of $R(\omega)\propto \omega^{-4}$ for $\omega\rightarrow\infty$
are used. The results are displayed in Fig.~\ref{optics}b and d
(note the different scales).

 In both directions the room-temperature
conductivity spectra consist of a narrow Drude-like component
\cite{DresselGruner02}
\begin{equation}
\sigma(\omega) = \frac{\omega_p^2\tau}{4\pi}\frac{1}{1+\omega^2\tau^2}
\end{equation}
and two broad maxima at higher frequencies. The Drude contribution
is described by the plasma frequency $\Omega_p=\omega_p/(2\pi
c)=3550$~\cm\  and the scattering rate $\Gamma=1/(2\pi
\tau)=180$~\cm\ for $E\bot$~stacks, and $\Omega_p=2300$~\cm\   and
$\Gamma=140$~\cm\  for the polarization $E\parallel$~stacks. An
intensive maximum lies at about 700~\cm, correspondingly a
pseudogap is observed in the spectra at 300~\cm. At higher
frequencies a broad and weak maximum can be seen at about
2600~\cm.

The spectral weight of the optical conductivity \cite{DresselGruner02}
\begin{equation}
\frac{\omega_p^2}{8}=\int\sigma(\omega) {\rm d}\omega
\end{equation}
shifts to low frequencies when the temperature is reduced. Although the Drude
contribution becomes extremely narrow,  its plasma frequency increases slightly;
at
$T=10$~K the corresponding fit yields $\Omega_p = 4050$~\cm\  and
$\Gamma=30$~\cm\
for $E\bot$~stacks, and  $\Omega_p=4050$~\cm\  and $\Gamma=45$~\cm\ for
$E\parallel$~stacks. This temperature behavior is quite common for organic
conductors and is
observed at the $\kappa$-phase of BEDT-TTF, for instance.\cite{Kornelsen89}

The 700~\cm\ peak considerably shifts to low frequencies as the temperature
decreases
and lies at about 350~\cm\
at $T=100$~K; thus the low-frequency gap becomes less pronounced.  The described
behavior is basically the same for both polarizations of  the electric field,
except
that for $E$ perpendicular to the stacks the spectral weight of this maximum
grows by a factor of 1.5
when going from $T=300$~K to 100~K while in the direction parallel to the stacks
the
intensity of this maximum does not change.

\begin{figure}
\includegraphics*[width=6cm]{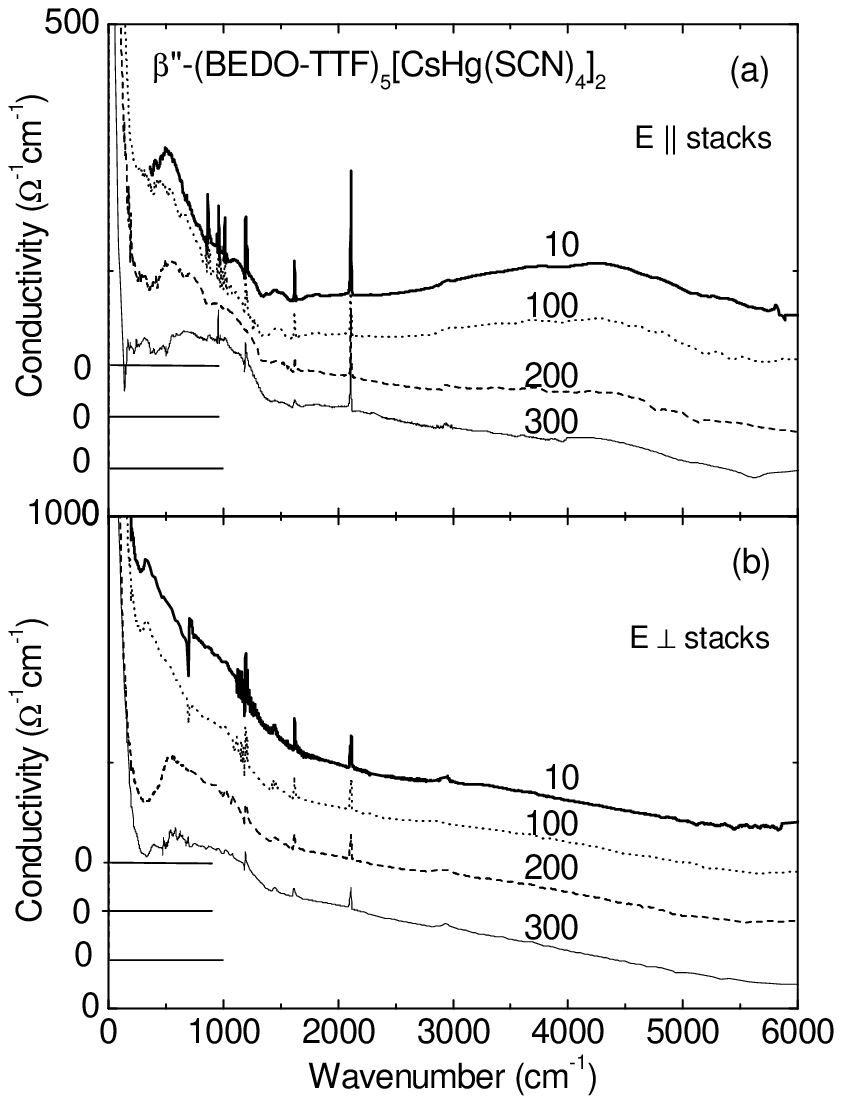}
\caption{\label{optics2}
Optical conductivity spectra of $\beta^{\prime\prime}$-(BEDO-
TTF)$_5$[CsHg(SCN)$_4$]$_2$
for the polarizations (a) E$\parallel$stacks and (b) E$\bot$stacks for different
temperatures as indicated. The curves are shifted for clarity reasons.
}
\end{figure}
An in-plane anisotropy observed in dc and microwave resistivity
measurements is also clearly seen in the different temperature
behavior of $\sigma(\omega)$ below 100~K. The spectra for
$E\bot$~stacks change only slightly between 100 and 10~K, in
agreement with the constant microwave resistivity in this
temperature range; in particular no changes are noted in the
mid-infrared range. Parallel to the stacks a wide band around
4000~\cm\ appears at 100~K, as is clearly seen in
Fig.~\ref{optics2}a where the conductivity spectra of different
temperatures are off-set. Simultaneously, the intensity of optical
conductivity $\sigma(\omega)$ in the mid-infrared region between
1000-3000~\cm\ decreases; also the low-frequency maximum becomes
smaller. This corresponds to the abrupt increase of the the
microwave resistivity at 100~K. Although $\rho(T)$ becomes
metallic below 30~K, the 4~K resistivity is still higher than that
at 100~K, in accordance with our optical results: the 4000~\cm\ is
present at the lowest measured temperature of 10~K.

\section{Discussion}

The most important observations of our optical experiments on
$\beta^{\prime\prime}$-(BEDO-TTF)$_5$\-[CsHg\-(SCN)$_4$]$_2$  are
an extremely narrow Drude-component and a gap at about 300~\cm. A
similar overall behavior was previously found in BEDT-TTF bases
organic conductors with a metallic ground
state.\cite{Dressel03,Haas00} In particular the fact that the
low-frequency spectral weight increases several times when the
temperature decreases from $T=300$~K to helium temperature, as
observed for quite a number of them,
\cite{Kornelsen89,Dressel03,Haas00,Eldridge91b,Jacobsen85,Jacobsen87}
cannot be simply explained by a reduced
scattering rate which leaves the spectral weight
conserved.\cite{DresselGruner02} Therefore these effects are
ascribed to the influence of electron-electron correlations.

It is interesting to compare our findings on
$\beta^{\prime\prime}$-(BEDO-TTF)$_5$\-[CsHg(SCN)$_4$]$_2$ with
the optical results \cite{dong1999} obtained on a superconductor
$\beta^{\prime\prime}$-(BEDT-TTF)$_2$\-SF$_5$\-CH$_2$CF$_2$SO$_3$
(T$_c$=5.2~K) which is a 1/4-filled compound with a similar
packing of the cation layers. The reflectivity of
$\beta^{\prime\prime}$-(BEDT-TTF)$_2$SF$_5$CH$_2$CF$_2$SO$_3$ is
lower than that of
$\beta^{\prime\prime}$-(BEDO-TTF)$_5$\-[CsHg(SCN)$_4$]$_2$,
although the dc conductivity is of the same order of magnitude.
Very similar to our optical data, the spectra of the
BEDT-TTF-based compound show two maxima in $\sigma(\omega)$ at
about 1000 and 2500~\cm; the low-frequency maximum grows and
shifts to lower frequencies with decreasing temperature, while no
Drude peak appears in the investigated temperature and frequency
range.\cite{dong1999}

The structural data on the two discussed
$\beta^{\prime\prime}$-salts of the BEDT-TTF and BEDO-TTF families
show that although the number of short contacts between the
neighboring molecules is larger in the case of
$\beta^{\prime\prime}$-(BEDO-TTF)$_5$\-[CsHg\-(SCN)$_4$]$_2$,
\cite{zhilya,Geiser96,Lyubovskii,Ward} the values calculated for
the intra-stack transfer integrals are equal. The inter-stack
overlap, on the other hand, is nearly twice larger for the
BEDT-TTF-based compound than for the BEDO-TTF one. If we assume
that the Coulomb interactions are comparable for both compounds,
the ratios of $U/t$ and $V/t$ are estimated to be larger for
$\beta^{\prime\prime}$-(BEDO-TTF)$_5$\-[CsHg(SCN)$_4$]$_2$
compared to
$\beta^{\prime\prime}$-(BEDT-TTF)$_2$\-SF$_5$\-CH$_2$\-CF$_2$\-SO$_3$.
Since the Coulomb effects do not drive the system insulating, but
we observe a Drude peak in our spectra (Fig.~\ref{optics}),
additional effects have to be considered.

An essential difference between these two
$\beta^{\prime\prime}$-salts is the filling of the  conduction
band which is 1/4 for the BEDT-TTF-compound and 1/5 for
$\beta^{\prime\prime}$-(BEDO-TTF)$_5$\-[CsHg\-(SCN)$_4$]$_2$. We
suggest  that this deviation from the quarter-filled band leads to
an appearance of the narrow Drude contribution, while the
low-frequency gap is still present in the spectra. The following
theoretical considerations confirm this idea.

A theoretical model of charge-ordering in layered molecular
crystals with quarter-filled conduction bands has been recently
put forward \cite{Seo00,McKenzie01} to describe this family of
$\beta^{\prime\prime}$-com\-pounds.\cite{Ward,Jones} Assuming a
simple square lattice to describe the quarter-filled molecular
planes of the crystal, a checkerboard charge-ordered state is
induced if values of the on-site $U$ and inter-site Coulomb
repulsion $V$ become large enough with respect to the
nearest-neighbors hopping integral $t$. Although for certain
ratios of $U/t$ and $V/t$ electronic correlations are not strong
enough to push the system into a checkerboard insulating state,
they nevertheless lead to strong modifications in the distribution
of the optical weight: for instance a low frequency peak appears
at the edge of a broad mid-infrared
band.\cite{McKenzie01,Dressel03} This is related to the dynamics
of quasiparticles interacting with short range checkerboard charge
fluctuations. For values of $V \gtrsim V_c \approx 2t$ and
sufficiently large $U=20t$, the Drude weight obtained from Lanczos
diagonalization is suppressed signalling the occurrence of a
metal-insulator transition (at quarter-filling).

This situation is changed as we dope the system away from one quarter filling.  Let us still model the
materials by means of an extended Hubbard model on a square lattice at arbitrary filling. A simple
understanding of the electronic properties of the system can be obtained by considering the case in which
we dope the insulating checkerboard charge-ordered state with only one hole as discussed earlier by Ohta
and collaborators.\cite{Ohta94} The single hole (which is an empty site in the checkerboard) has non-zero
probability (for $V > V_c=2t$ and $U>>t$), to move to a second-nearest neighbor in the diagonal direction
by two successive hopping processes through its nearest neighbors. This is of course a virtual second
order process because the electron has to hop twice along its nearest neighbors at the expense of some
energy. In a subsequent process the electron can hop to its third-nearest neighbors and so on. Hence,
quasiparticles can disperse with renormalized hopping along the lattice and the system is expected to be
always metallic (at a finite value of $V$). Hence, while at 1/4-filling a metal-insulator transition at a
finite $V_c$ is expected, this is not the case at 1/5-filling. This explains why the 1/5-filled salt
exhibits a robust Drude component in contrast to the 1/4-filled $\beta''$ salts. The short range
checkerboard charge fluctuations which lead to signatures in the dynamical properties of the 1/4-filled
system (such as the appearance of the low frequency peak in the optical conductivity) are also expected to
be present in the 1/5-filled system. However, these signatures are anticipated to be somewhat suppressed
at 1/5-filling with respect to 1/4-filling for a given value of $V$ because in the former case, the
checkerboard charge ordered state is modified by the presence of mobile empty sites. This arguments are
consistent with the observation that while the spectral weight of the feature appearing at low frequencies
in the optical conductivity increases by factor of 5 as the temperature is reduced from room temperature
to the lowest temperature at 1/4-filling, at 1/5-filling this feature is only enhanced by a
factor of approximately 1.5.\cite{footnote3}

\begin{figure}
\includegraphics*[width=7cm]{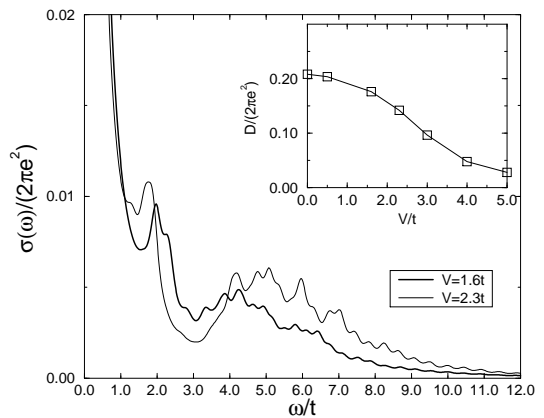}
\caption{\label{figdrude} Optical conductivity and Drude weight
from exact diagonalization calculations of an extended Hubbard
model on a square lattice.  The system is at 1/5-filling, $U=20t$
and the cluster size is $L=20$. A broadening of $\eta=0.2t$ has
been used for comparison to experimental data.  At 1/5-filling, the system
remains
metallic for any value of $V$ in contrast with the 1/4-filled
salt, which becomes insulating for $V \gtrsim V_c \approx 2t$.
These calculations are consistent with the strong Drude component
observed in the optical conductivity of 1/5-filled salts in
contrast to the weaker one found for their 1/4-filled
counterparts. Although strongly suppressed with respect to the
1/4-filled salt, a low frequency feature is evident for the plotted
values of $V/t$.}
\end{figure}
In Fig. \ref{figdrude} we show the behavior of the optical
conductivity and the Drude weight at 1/5-filling for different
values of $V$ an $U=20t$, computed from exact diagonalization of
an extended Hubbard model on a square lattice. We observe how, at
1/5-filling, the Drude weight remains finite; the system remaining
metallic for any value of $V$.  This is in contrast to the
metal-insulator transition found at $V \approx 2t$ for the
1/4-filled system.\cite{Calandra02} The optical conductivity for
typical values of $V/t$ displays a broad band and a feature at low
frequencies; much smaller than the one found for the 1/4-filled
case.\cite{Dressel03} The behavior of the optical spectra at low
frequencies when going from the 1/4 to the 1/5-filled system is
consistent with exact diagonalization results.\cite{footnote4}

While the features ascribed to the influence of electronic
correlations are observed in the spectra of both polarizations,
the non-metallic behavior of dc, microwave, and optical
conductivity is seen only for $E$ parallel to the stacks and can
be ascribed to structural effects. One of the major changes below
100~K affects the broad band at about 2500~\cm.

A wide band in the mid-infrared region is typical for the
conductivity spectra of most BEDT-TTF and BEDO-TTF based
materials. It can appear due to to structural peculiarities: for
example, when the filling of the conduction band is not equal to
one half, a dimerization in the stacks of the organic molecules
can cause the conduction band to split; transitions across this
gap result in a maximum in the optical conductivity
spectra.\cite{kuroda} Since in the cation layers of
$\beta^{\prime\prime}$-(BEDO-TTF)$_5$\-[CsHg(SCN)$_4$]$_2$ the
molecules are weakly modulated along the stacks already at room
temperature, the wide band with a maximum at about 2500~\cm\ can
be assigned to transitions across the gap caused by this
structural modulation. The shift of this band up to 4000~\cm\ and
the fact that its intensity for the polarization parallel to the stacks grows at
temperatures below 150~K points towards a strengthening of the
modulation. The spectral weight of this band increases on the
expense of the low-frequency feature: below $T\approx 100$~K the
intensity of the 350~\cm\ maximum is reduced and the 250~\cm\ gap
becomes more distinct; consequently, the dc and microwave resistivity
increase.

Most changes in structure and charge on the BEDT-TTF molecules in
the organic conductors can be nicely followed by observing changes
of  EMV-coupled vibrational features~\cite{Yartsev}. The  $a_g$
vibrations of central C=C bonds of the BEDO-TTF molecule
observed\cite{we} at 1619~\cm\ is known to be most sensitive to
the changes of charge on the molecule.\cite{Drozdova} Its position
does not show any variations with temperature, which suggests that
the charge on the BEDO-TTF molecules does not change considerably.
On the other hand, for the polarization parallel to the stacks two
$a_g$ vibrational modes at 864 and 1015~\cm\ appear simultaneously
with the growth of the 4000~\cm\ electronic band.\cite{we} The
appearance of new vibrational features indicates a lowering of
symmetry: a symmetry center disappears. Since $a_g$ features
couple to electrons which are localized on clusters of molecules,
we can conclude that these changes in the spectra occur due to a
strengthening of the modulation within the stacks of BEDO-TTF
molecules, which leads to the appearance of clusters (e.g.\ dimers
of BEDO-TTF) and lowers the symmetry.\cite{footnote2} However, for
1/5-filled system the most primitive modulations (dimerization,
teramerisation, etc.) will not cause a gap at the Fermi-surface
and lead to a insulating behavior. Maybe this is the reason why we
observe such a complicated behavior in the dc and microwave
resistivity instead of a simple metal-to-insulator transition.
Interestingly, this non-metallic behavior is absent in the
direction of the larger overlap of the neighboring molecule
orbitals, i.e.\ for the polarization $E$ perpendicular to the stacks.

\section{Conclusions}
The present optical study of
$\beta^{\prime\prime}$-(BEDO-TTF)$_5$\-[CsHg\-(SCN)$_4$]$_2$
 in the infrared range between 60 and 6000~\cm\
from room temperature down to 10~K allows us to investigate and
distinguish between two effects relevant to most of the
two-dimensional organic conductors. Electronic correlations lead
to the presence of a narrow Drude-like contribution and to a
pseudogap at about 300~\cm\ in the optical conductivity
$\sigma(\omega)$. These features are observed for both principal
directions in the conducting plane, parallel and perpendicular to
the stacks, and in the whole temperature range, from 10~K  up to
room temperature. The maximum in the optical conductivity due to
excitations across the pseudogap shift to low frequencies as the
temperature decreases. Comparing our results on the BEDOT-TTF
systems with the optical data of BEDT-TTF at 1/4-filling, we
propose that the robustness of the Drude contribution in our
spectra is due to the fact that the conduction band is 1/5 filled.
We have used an extended Hubbard model with $U>>t$ on a square
lattice at different fillings to model the molecular planes. At
quarter-filling there is a critical value, $V_c \approx 2t$, at
which a metal-insulator transition to a checkerboard charge
ordered state occurs. Optical conductivity close to but in the
metallic side of this transition displays a strong suppression of
the Drude component and transfer of weight to a broad mid-infrared
band. At 1/5-filling, however, empty sites induced by doping the
checkerboard propagate along the molecular planes leading to
metallic behavior. This explains the strong Drude component
observed in the optical conductivity at 1/5-filling. It would very
interesting to analyze materials with other fillings so that an
understanding of the dynamics of holes in a charge ordered
background can be achieved.

At temperatures below 100~K structural modulations along the
stacks of BEDO-TTF molecules activate a band at 4000~\cm\ and new
EMV-coupled features for the polarization $E$ parallel to the stacks. In
the dc and microwave resistivity this effect is observed as a
non-metallic behavior in this direction.

\begin{acknowledgments}
We are grateful to R. B. Lyubovskii for d.c.resistivity
measurements. We thank R. Vlasova and V. Semkin for fruitful
discussions. N.D. was supported by ``Leading scientific school''-
2200.2003.2, by the Deut\-sche Akademische Auslandsdienst (DAAD),
and by the Alexander von Humboldt-Foundation. O.A.B., E.I.Z and
R.N.L. acknowledge support of RFBR 01-03-33009 grant. J.M. has
been supported through a Marie Curie Fellowship of the European
Community Program ``Improving Human Potential'' under contract No.
HPMF-CT-2000-00870.
\end{acknowledgments}


\begin{thebibliography}{99}
\bibitem{footnote}These are results from measurements performed with a cooling
rate of 1~K/min;
for faster cooling the non-monotonic behavior found parallel to the stacks is
less
pronounced.
\bibitem{footnote2}We cannot distinguish between a dimerization or some
other sort of structural modulation.
\bibitem{footnote3}The idea of superconductivity droven by charge-ordering fluctuations
agrees with the fact that the quarter-filled salt
$\beta^{\prime\prime}$-(BEDT-TTF)$_2$\-SF$_5$\-CH$_2$CF$_2$SO$_3$
becomes superconducting on cooling below  $T_c=5.2$~K, while the
1/5-filled
$\beta^{\prime\prime}$-(BEDO-TTF)$_5$\-[CsHg\-(SCN)$_4$]$_2$
remains a bad metal at any temperature.
\bibitem{footnote4}The spectral weight of the Drude component in
the 1/5-filled
$\beta^{\prime\prime}$-(BEDO-TTF)$_5$\-[CsHg\-(SCN)$_4$]$_2$ is
about one order of magnitude higher than in 1/4 filled
$\alpha$-ET$_2$KHg(SCN)$_4$~\cite{Dressel03}.
\end{thebibliography}
\end{document}